\documentclass{aa}
\usepackage{graphicx}
\usepackage{txfonts}
\usepackage{tabularx}
\usepackage{multirow}
\usepackage{amssymb}
\usepackage{latexsym}

\begin{document}

\title{Vacuum-UV spectroscopy of interstellar ice analogs.} 
\subtitle{I. Absorption cross-sections of polar-ice molecules.}
\author{G. A. Cruz-Diaz \inst{1}, G. M. Mu\~{n}oz Caro \inst{1}, Y.-J. Chen \inst{2,} \inst{3}, and T.-S. Yih \inst{3}}
\offprints{Gustavo A. Cruz-Diaz}
\institute{Centro de Astrobiolog\'{\i}a, INTA-CSIC, Carretera de Ajalvir, 
km 4, Torrej\'on de Ardoz, 28850 Madrid, Spain\\
email: cruzdga@cab.inta-csic.es, munozcg@cab.inta-csic.es 
\and Space Sciences Center and Depeartment of Physics and Astronomy, University of Southern California, Los Angeles, CA 90089-1341, USA
\and Department of Physics, National Central University, Jhongli City, Taoyuan Country 32054, Taiwan}
\date{Received - , 0000; Accepted - , 0000}
\authorrunning{G. A. Cruz-Diaz et al.}  
\titlerunning{Vacuum-UV spectroscopy of interstellar ice analogs}

\abstract
{The VUV absorption cross sections of most molecular solids present in interstellar ice mantles with the exception of 
H$_{2}$O, NH$_{3}$, and CO$_{2}$ have not been reported yet. Models of ice photoprocessing depend on the VUV absorption cross 
section of the ice to estimate the penetration depth and radiation dose, and in the past, gas phase cross section values were used as 
an approximation.}
{We aim to estimate the VUV absorption cross section of molecular ice components.}
{Pure ices composed of CO, H$_{2}$O, CH$_{3}$OH, NH$_{3}$, or H$_{2}$S were deposited at 8 K. The column density of the ice 
samples was measured \emph{in situ} by infrared spectroscopy in transmittance. VUV spectra of the ice samples were collected 
in the 120-160 nm (10.33-7.74 eV) range using a commercial microwave-discharged hydrogen flow lamp.}
{We provide VUV absorption cross sections of the reported molecular ices. Our results agree with those previously 
reported for H$_2$O and NH$_3$ ices. Vacuum-UV absorption cross section of CH$_3$OH, CO, and H$_2$S in solid phase are reported for the first time. 
H$_2$S presents the highest absorption in the 120-160 nm range.}
{Our method allows fast and readily available VUV spectroscopy of ices without the need to use a synchrotron beamline. We found that the ice absorption 
cross sections can be very different from the gas-phase values, and therefore, our data will significantly improve models that simulate the VUV photoprocessing and 
photodesorption of ice mantles. Photodesorption rates of pure ices, expressed in molecules per absorbed photon, can be derived from our data.}

\maketitle

\label{Intro}
\section{Introduction}

Ice mantles in dense cloud interiors and cold circumstellar environments are composed mainly of H$_{2}$O and other species such as CO$_2$, CO, CH$_4$, CH$_{3}$OH, and NH$_{3}$ 
(Mumma \& Charnley 2011 and ref. therein). Some species with no permanent or induced dipole moment such as O$_2$ and N$_2$, cannot be easily observed in the infrared, but are also 
expected to be present in the solid phase (e.g., Ehrenfreund \& van Dishoeck 1998). The relative to water abundances of the polar 
species CO, CH$_{3}$OH, and NH$_{3}$ are between 0-100\%, 1-30\%, and 2-15\%. In comets the abundances 
are 0.4-30\%, 0.2-7\%, 0.2-7\%, and $\sim$ 0.12-1.4\% for CO, CH$_3$OH, NH$_3$, and H$_2$S 
(Mumma \& Charnley 2011 and ref. therein). We therefore included H$_{2}$S in this study because this reduced species was 
detected in comets and is expected to form in ice mantles (Jim\'enez-Escobar \& Mu\~{n}oz Caro 2011).
In the coldest regions where ice mantles form, thermally induced reactions are inhibited. Therefore, irradiation processes by UV-photons or cosmic rays may play an important 
role in the formation of new species in the ice and contribute to the desorption of ice species to the gas phase. Cosmic rays penetrate deeper into the cloud interior than UV-photons, 
generating secondary UV-photons by excitation of H$_2$ molecules. This secondary UV-field will interact more intensively with the grain mantles than direct impact 
by cosmic rays (Cecchi-Pestellini \& Aiello 1992, Chen et al. 2004). VUV photons have the power to excite or dissociate molecules, leading to a complex 
chemistry in the grains. 
The VUV-region encloses wavelengths from 200 nm to about 100 nm, since the term extreme-ultraviolet (EUV) is often used for shorter 
wavelengths.

The estimation of the VUV-absorption cross sections of molecular ice components allows one to calculate the photon absorption of icy grains in that range. 
In addition, the VUV-absorption spectrum as a function of photon wavelength is required to study the photodesorption processes over the full photon emission energy range. 
Photoabsorption cross-section measurements in the VUV-region were 
performed for many gas phase molecules. Results for small molecules in the gas phase were summarized by Okabe \cite{Okabe}, but most of these cross-section 
measurements for molecular bands with fine structure are severely distorted by the instrumental bandwidths (Hudson \& Carter 1968). 
The integrated cross sections are less affected by the instrumental widths, and approach the true cross sections as 
optical depth approaches zero. Therefore, the true integrated cross section can be obtained from series of data taken with different 
column densities (Samson \& Ederer 2000). 

The lack of cross-section measurements in solids has led to the assumption that the cross sections of molecules in ice mantles were similar to the gas phase values. 
In recent years the VUV-absorption spectra of solid H$_{2}$O, CH$_{3}$OH, NH$_{3}$, CO$_2$, 
O$_2$, N$_2$, and CH$_4$ were reported by Mason et al. \cite{Mason}, Kuo et al. \cite{Kuo}, and Lu et al. \cite{Lu2}. But the 
VUV-absorption cross sections were only estimated for solid H$_{2}$O, NH$_{3}$, and CO$_{2}$ (Mason et al. 2006). Furthermore, all previous works have been performed 
using synchrotron monochromatic light as the VUV-source, 
scanning the measured spectral range.

In the present work, we aim to provide accurate measurements of the VUV-absorption cross sections of interstellar ice polar 
components including H$_{2}$O, CH$_{3}$OH, NH$_{3}$, CO, and H$_{2}$S. The use of a hydrogen VUV-lamp, commonly used in ice irradiation experiments, 
limits the spectroscopy to the emission range between 120 and 160 nm, but this has several advantages: the measurements are easy to perform and can be made regularly in the laboratory, 
without the need to use synchrotron beam time. A second paper will be dedicated to the nonpolar molecular ice components including CO$_2$, CH$_4$, N$_2$, and O$_2$. 
In Sect. ~\ref{Expe} the experimental protocol is described. Sect. ~\ref{VUV} provides the determination of VUV-absorption cross 
sections for the different ice samples. The astrophysical implications are presented in Sect. ~\ref{Astro}. The conclusions are 
summarized in Sect. ~\ref{Conclusions}.

\section{Experimental protocol}
\label{Expe}

\begin{figure}[ht!]
\centering
\includegraphics[width=8.5cm]{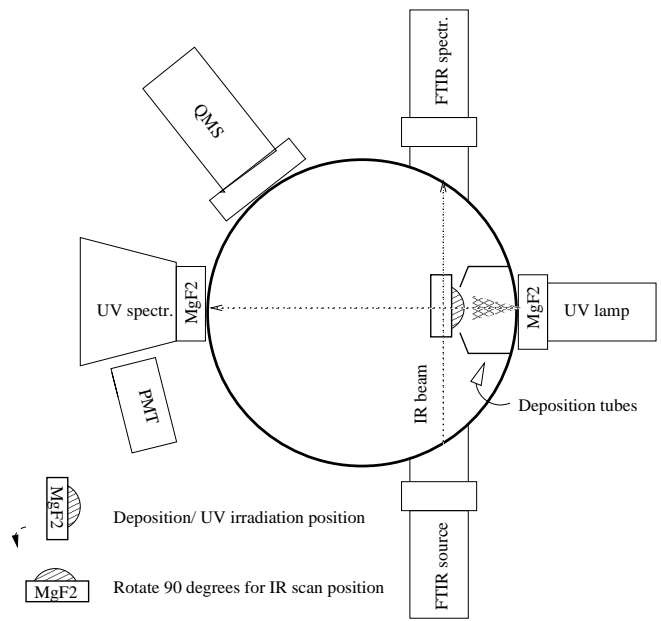}
\caption{Scheme of the main chamber of ISAC. The VUV-light intersects three MgF$_{2}$ windows before it enters the UV-spectrometer, 
but the emission spectrum that the ice experiences corresponds to only one MgF$_{2}$ window absorption, the one between the VUV-lamp 
and the ISAC main chamber. FTIR denotes the source and the detector used to perform infrared spectroscopy. QMS is the quadrupole mass spectrometer 
used to detect gas-phase species. PMT is the photomultiplier tube that makes the ultraviolet spectroscopy posible.}
\label{setup}
\end{figure}

The experiments were performed using the interstellar astrochemistry chamber (ISAC), see Fig. ~\ref{setup}. This set-up and the 
standard experimental protocol were described in Mu\~noz Caro et al. \cite{Caro1}. ISAC mainly consists of an ultra-high-vacuum (UHV) 
chamber, with pressure typically in the range P = 3-4.0 $\times$ 10$^{-11}$ mbar, where an ice layer made by deposition of a gas species 
onto a cold finger at 8 K, achieved by means of a closed-cycle helium cryostat, can be UV-irradiated. The evolution of the solid sample 
was monitored with \emph{in situ} transmittance FTIR spectroscopy and VUV-spectroscopy. The chemical components used for the 
experiments described in this paper were H$_2$O(liquid), triply distilled; CH$_3$OH(liquid), Panreac Qu\'{\i}mica S. A. 99.9\%; CO(gas), Praxair 99.998\%; 
NH$_3$(gas), Praxair 99.999\%; and H$_{2}$S(gas), Praxair 99.8\%. The deposited ice layer was photoprocessed with an F-type microwave-discharged hydrogen flow 
lamp (MDHL), from Opthos Instruments. The source has a VUV-flux of $\approx$ 2 $\times$ 10$^{14}$ cm$^{-2}$ s$^{-1}$ at the sample 
position, measured by CO$_{2}$ $\to$ CO actinometry, see Mu\~noz Caro et al. \cite{Caro1}. The Evenson cavity of the lamp is refrigerated 
with air. The VUV-spectrum is measured routinely {\em in situ} during the irradiation experiments with the use of a McPherson 0.2-meter 
focal length VUV monochromator (model 234/302) with a photomultiplier tube (PMT) detector equipped with a sodium salicylate window, 
optimized to operate from 100-500 nm (11.27-2.47 eV), with a resolution of 0.4 nm. The characterization of the MDHL spectrum has been  
studied before by Chen et al. \cite{Asper1} and will be discussed in more detail by Chen et al. \cite{Asper2}.

The interface between the MDHL and the vacuum chamber is a MgF$_{2}$ window. The monochromator is located at the rear end 
of the chamber, separated by another MgF$_{2}$ window. This means that the measured background spectra, that is without an ice sample 
intersecting the VUV-light cone, are the result of the radiation that intersects two MgF$_{2}$ windows.

Grating corrections were made for the VUV-spectra in the range of 110-180 nm (11.27-6.88 eV). 
The mean photon energy was calculated for the spectrum corresponding to only one MgF$_{2}$ window intersecting the VUV-lamp emission, 
since that is the mean photon energy that the ice sample experiences. This one-window spectrum, displayed in Fig. 2, 
was measured by directly coupling the VUV-lamp to the spectrometer with a MgF$_{2}$ window acting as the interface. The proportion 
of Ly-$\alpha$ in 110-180 nm range is 5.8\%, lower than the 8.4\% value estimated by Chen et al. \cite{Asper2}. This difference 
could be due to the lower transmittance of the MgF$_{2}$ window used as interface between the MDHL and the UHV-chamber, 
although the different position of the pressure gauge measuring the H$_{2}$ flow in the MDHL may also play a role.

\begin{figure}[ht!]
\centering
\includegraphics[width=\columnwidth]{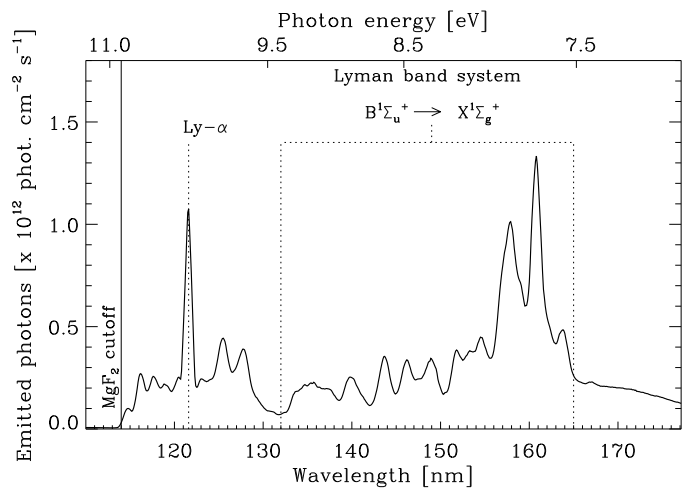}
\caption{UV-photon flux as a function of wavelength of the MDHL in the 110 to 170 nm range estimated with the total photon flux calculated using actinometry. The spectrum 
corresponds to a measurement with one MgF$_{2}$ window intersecting the emitted VUV-light cone. This spectrum is the one experienced by the ice sample. The VUV-emission is 
dominated by the Ly-$\alpha$ peak (121.6 nm) and the Lyman band system.}
\label{Uno}
\end{figure}

It was observed that most of the energy emitted by the VUV-lamp lies below 183 nm (6.77 eV) and the MgF$_{2}$ window cutoff occurs at 114 nm 
(10.87 eV). The mean photon energy in the 114-180 nm (10.87-6.88 eV) range is E$_{photon} = 8.6$ eV. The main emission bands are 
Ly-$\alpha$ at 121.6 nm (10.20 eV) and the molecular hydrogen bands centered on 157.8 nm (7.85 eV) and 160.8 nm (7.71 eV) for a 
hydrogen pressure of 0.4 mbar, see Fig.~\ref{Uno}.

\section{VUV-absorption cross section of interstellar polar ice analogs}
\label{VUV}

We recorded VUV-absorption spectra of pure ices composed of CO, H$_{2}$O, CH$_{3}$OH, NH$_{3}$, and H$_{2}$S. 
For each ice spectrum a series of three measurements was performed: i) the emission spectrum of the VUV-lamp was measured to 
monitor the intensity of the main emission bands, ii) the emission spectrum transmitted by the MgF$_{2}$ substrate window was measured 
to monitor its transmittance, and iii) the emission spectrum transmitted by the substrate window with  
the deposited ice on top was measured. The absorption spectrum of the ice corresponds to the spectrum of the substrate with the ice 
after subtracting the bare MgF$_2$ substrate spectrum. In addition, the column density of the ice sample was measured by FTIR in 
transmittance. The VUV-spectrum and the column density of the ice were therefore monitored in a single experiment for 
the same ice sample. This improvement allowed us to estimate the VUV-absorption cross section of the ice more accurately. The column 
density of the deposited ice obtained by FTIR was calculated according to the formula
\begin{equation}
\hspace{3cm}
N= \frac{1}{\mathcal{A}} \int_{band} \tau_{\nu}d{\nu},
\label{1}
\end{equation}
where $N$ is the column density of the ice, $\tau_{\nu}$ the optical depth of the band, $d\nu$ the wavenumber differential, 
in cm$^{-1}$, and $\mathcal{A}$ is the band strength in cm molecule$^{-1}$. The integrated absorbance is equal to 
0.43 $\times$ $\tau$, where $\tau$ is the integrated optical depth of the band. The VUV-absorption cross section was estimated according to the Beer-Lambert law,
\begin{eqnarray}
\hspace{3cm}
I_t(\lambda) &=& I_0(\lambda) {e}^{-\sigma(\lambda) N} \nonumber \\
\sigma(\lambda) &=& - \frac{1}{N} \ln \left( \frac{I_t(\lambda)}{I_0(\lambda)} \right) \nonumber \\ 
\text{with} \quad N &\approx& \frac{N_i + N_f}{2}
\label{2}
\end{eqnarray}
where I$_{t}(\lambda)$ is the transmitted intensity for a given wavelength $\lambda$, I$_{0}(\lambda)$ the incident intensity, $N$ is an average ice
column density before ($N_i$) and after ($N_f$) irradiation in cm$^{-2}$, and $\sigma$ is the cross section in cm$^{2}$. 
It is important to notice that the total VUV-flux value emitted by the lamp does not affect the VUV-absorption spectrum of the ice sample, since it is obtained by substracting 
two spectra to obtain the absorbance in the VUV. 

A priori, the VUV-absorption cross section of the ice was not known. Therefore, several measurements for different values of the ice 
column density were performed to improve the spectroscopy. Table ~\ref{table1} provides the infrared-band position and the band strength 
used to obtain the column density of each ice component, along with the molecular dipole moment, since the latter has an effect on 
the intermolecular forces that operate in the ice that distinguish solid spectroscopy from gas phase spectroscopy in the IR and VUV ranges. 
The gas-phase cross sections published in the literature were also represented for comparison.
We note that the majority 
of the gas-phase cross sections were performed at room temperature. The gas-phase spectrum can vary significantly when the gas sample 
is cooled to cryogenic temperatures, but it will still differ from the solid-phase spectrum, see Yoshino et al. \cite{Yoshino} and 
Cheng et al. \cite{Cheng2}. 

\begin{table}
\centering
\caption{Infrared-band positions and strengths ($\mathcal{A}$), deposited column density ($N$ in 10$^{15}$ molec./cm$^{2}$) and 
molecular dipole moment ($\mu$) of the samples used in this work. Error estimation in the column density is the sum of the error by the band 
strength, the error by thickness loss for photodesorption, and the MDHL, PMT, and multimeter stabilities.}
\tiny
\begin{tabular}{ccccc}
\hline
\hline
Species&Position&$\mathcal{A}$&$N$&$\mu$\\
&[cm$^{-1}$]&[cm/molec.]&[$\times$ 10$^{15}$ molec./cm$^{2}$)]&[D]\\
\hline
CO& 2139 & 1.1 $\pm$ 0.1 $\times10^{-17 \; a}$ &207$^{+16}_{-16}$ &0.11\\
&&&&\\
H$_{2}$O& 3259 & 1.7 $\pm$ 0.2 $\times10^{-16 \; b}$ &91$^{+7}_{-7}$ &1.85\\
&&&&\\
CH$_{3}$OH& 1026 & 1.8 $\pm$ 0.1 $\times10^{-17 \; c}$ &62$^{+5}_{-9}$ &1.66\\
&&&&\\
NH$_{3}$& 1067 & 1.7 $\pm$ 0.1  $\times10^{-17 \; c}$ &67$^{+5}_{-8}$ &1.47\\
&&&&\\
H$_{2}$S& 2544 & 2.0 $\pm$ 0.2 $\times10^{-17 \; d}$ &37$^{+3}_{-9}$ &0.95\\
\hline
\end{tabular}
\\
$^a$Jiang et al. 1975, $^b$calculated for this work, $^c$d'Hendecourt \& Allamandola 1986, $^d$Jim\'enez-Escobar \& Mu\~noz Caro 2011. \\
\label{table1}
\end{table}

The performance of VUV-spectroscopy inevitably leads to exposure of the sample to irradiation, often causing
photodestruction or photodesorption of the ice molecules to some extent during spectral acquisition 
(e.g., d'Hendecourt et al. 1985, 1986; Bernstein et al. 1995; Gerakines et al. 1995, 1996; Schutte 1996; 
Sandford 1996; Mu\~noz Caro et al. 2003, 2010; Dartois 2009; \"Oberg et al. 2009b,c,d; Fillion et al. 2012; Fayolle et al. 2011, 2013,  and ref. therein). Photoproduction of new 
compounds and the decrease of the starting ice column density was monitored by IR spectroscopy. Photodesorption was less intense for the reported species except for CO, in line with 
the previous works mentioned above; it will also contribute to the decrease of the column density of the ice during irradiation. Photoproduct formation of new compounds is different 
for each molecule. No photoproducts were detected after the 9 min VUV exposure required for VUV-spectroscopy, except for CH$_3$OH ice, where product formation was 2\%, 6\%, 7\%, and 
8\% for CO$_2$, CH$_4$, CO, and H$_2$CO relative to the initial CH$_3$OH column density estimated by IR spectroscopy. But the VUV-absorption spectrum of CH$_3$OH 
was not significantly affected, cf Kuo et al. \cite{Kuo}. Indeed, we show below that the discrete bands of the CO photoproduced in the CH$_3$OH ice matrix, with an intrinsic 
absorption higher than methanol in the VUV, are absent from the VUV-absorption spectrum.

Error values for the column density in Table ~\ref{table1} result mainly from the selection of the baseline for integration of the IR absorption band and the decrease of the 
ice column density due to VUV-irradiation during spectral acquisition. The band strengths were adapted from the literature, and their error estimates are no more than 10 \% of the 
reported values (Richey \& Gerakines 2012). The solid H$_2$O band strength in Table \ref{table1} was estimated by us using interference fringes in the infrared spectrum for a 
density of 0.94 g cm$^{-3}$ and a refractive index of 1.3. This estimation gives a value of 1.7 $\pm$ 0.2 $\times10^{-16}$ cm molec$^{-1}$, different from the Hagen et al. (1985) 
value (2 $\times10^{-16}$ cm molec$^{-1}$). The errors in the column density determined by IR spectroscopy were 16\%, 15\%, 23\%, 19 \%, and 32\% for solid CO, H$_2$O, 
CH$_3$OH, NH$_3$, and H$_2$S ices. The MDHL, photomultiplier tube (PMT), and multimeter stabilities lead to an estimated error of about 6\% in the values of the 
VUV-absorption cross sections of the ices. The largest error corresponds to H$_2$S ice, the most VUV photoactive molecule studied in this work. It presents a high 
photodestruction (Jim\'enez-Escobar \& Mu\~{n}oz Caro 2011), which leads to a fast decrease in its IR feature. 
Therefore, VUV-absorption cross-section errors result from the error values estimated above, using the expression

\begin{equation}
\delta(N) =  \sqrt{\frac{\delta_i^2 + \delta_j^2 + \delta_k^2 + ... + \delta_n^2}{n-1}} .
\end{equation}

The VUV-absorption cross-section spectra of CO, H$_2$O, CH$_3$OH, and NH$_3$ ices were fitted with Gaussian profiles using the band positions reported in the literature (Lu et al. 
2005, 2008; Kuo et al. 2007) as a starting point, see the red traces in Figs.~\ref{CO},~\ref{H2O},~\ref{CH3OH}, and ~\ref{NH3}. Table~\ref{tableGauss} summarizes 
the Gaussian profile parameters used to fit the spectra of these ices, deposited at 8 K. H$_{2}$S ice displays only a slightly decreasing absorption at longer wavelengths, and 
Gaussian deconvolution is thus not pertinent. Gaussian fits of the reported molecules were made with an in-house IDL code. The fits reproduce the VUV-absorption 
cross-section spectra well.

\begin{table}[ht!]
\centering
\caption{Parameter values used to fit the spectra of Gaussian profiles of the different molecular pure ices deposited at 8 K.}
\tiny
\begin{tabular}{cccc}
\hline
\hline
Molecule&Center&FWHM&Area\\
&[nm]&[nm]&[$\times$ 10$^{-17}$ cm$^{2}$ nm]\\
\hline
&&&\\
H$_{2}$O&120.0&22.1&12.0\\
&143.6&17.4&10.8\\
&154.8&8.0&0.9\\
&&&\\
CO&127.0&1.27&0.05\\
&128.9&1.27&0.07\\
&130.9&1.32&0.14\\
&132.8&1.27&0.24\\
&135.2&1.46&0.58\\
&137.6&1.41&0.95\\
&140.0&1.62&1.61\\
&142.8&1.65&1.98\\
&145.8&2.17&3.16\\
&148.8&2.24&3.45\\
&149.7&0.92&0.50\\
&152.0&2.12&3.09\\
&153.3&1.34&1.53\\
&156.3&2.24&3.10\\
&157.0&1.18&0.56\\
&&&\\
CH$_{3}$OH&118.0&30.6&27.7\\
&146.6&28.3&16.8\\
&&&\\
NH$_{3}$&121.2&34.1&33.8\\
&166.6&31.8&14.2\\
\hline
\end{tabular}
\label{tableGauss}
\end{table}

\subsection{Solid carbon monoxide}

The ground state of CO is X$^{1}\Sigma^{+}$ and its bond energy is E$_{b}$(C--O) = 11.09 eV (Okabe 1978). 

\begin{figure}[ht!]
\centering
\includegraphics[width=\columnwidth]{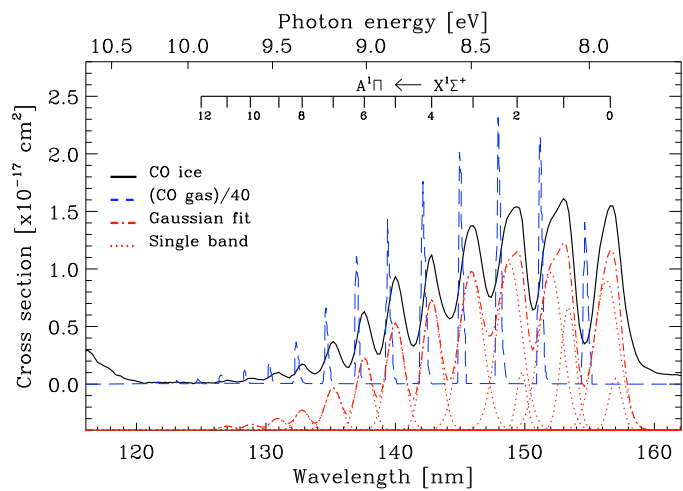}
\caption{VUV-absorption cross section as a function of photon wavelength (bottom X-axis) and VUV-photon energy (top X-axis) of pure CO 
ice deposited at 8 K, black solid trace. The blue dashed trace is the VUV-absorption cross-section spectrum of CO gas (divided by a factor of 40 to  
compare it with CO ice) adapted from Lee \& Guest \cite{Lee}. The fit, red dashed-dotted trace, is the sum of 15 Gaussians, dotted trace. It has been 
vertically offset for clarity. }
\label{CO}
\end{figure}

Fig.~\ref{CO} displays the CO fourth positive band system; it is attributed to the A$^{1}\Pi \leftarrow$ X$^{1}\Sigma^{+}$ 
system. Table ~\ref{TableCO} summarizes the transition, band position, and area of each feature at 8 K. We were able to observe twelve bands 
identified as (0,0) to (11,0) where the (0,0), (1,0), and (2,0) bands present a Davydov splitting, in agreement with Mason 
et al. \cite{Mason}, Lu et al. \cite{Lu}, and  Brith \& Schnepp \cite{Brith}. Like Mason et al. \cite{Mason}, we were unable to observe the (12,0) transition reported by 
Brith \& Schnepp \cite{Brith} and Lu et al. \cite{Lu}, 
because of the low intensity of this feature. We detected part of the transitions to the excited Rydberg states, 
B$^{1}\Sigma^{+}$, C$^{1}\Sigma^{+}$, and E$^{1}\Pi$ measured by Lu et al. \cite{Lu} as a broad band in the 116-121 nm (10.68-10.25 eV) 
region, despite the decreasing VUV-flux in this region. 

The average VUV-absorption cross section has a value of 4.7$^{+0.4}_{-0.4}$ $\times$ 10$^{-18}$ cm$^{2}$. This value is not very different from the one roughly estimated by 
Mu\~noz Caro et al. \cite{Caro1}, 3.8 $\times$ 10$^{-18}$ cm$^{2}$ in the 115-170 nm range based on Lu et al. \cite{Lu}. 
The total integrated VUV-absorption cross section has a value of 1.5$^{+0.1}_{-0.1}$ $\times$ 10$^{-16}$ cm$^{2}$ nm (7.9$^{+0.6}_{-0.6}$ $\times$ 10$^{-18}$ cm$^{2}$ 
eV) in the 116-163 nm (10.68-7.60 eV) spectral region. 
The VUV-absorption of CO ice at 121.6 nm is very low, an upper limit of $\leq$ 1.1$^{+0.8}_{-0.8}$ $\times$ 10$^{-19}$ cm$^{2}$ was estimated, while at 157.8 and 160.8 nm the 
VUV-absorption cross sections are 
6.6$^{+0.5}_{-0.5}$ $\times$ 10$^{-18}$ cm$^{2}$ and 0.9$^{+0.1}_{-0.1}$ $\times$ 10$^{-18}$ cm$^{2}$. The VUV-absorption spectrum of solid CO presents a maximum at 153.0 nm 
(8.10 eV) with a value of 1.6$^{+0.1}_{-0.1}$ $\times$ 10$^{-17}$ cm$^{2}$. 
Table ~\ref{TableCOgas} summarizes the transition, band position, and area of each feature present in the CO gas-phase spectrum (blue trace 
in Fig.~\ref{CO}) adapted from Lee \& Guest \cite{Lee}. The VUV-absorption cross section of CO in the gas phase has an average value of 20.7 $\times$ 10$^{-18}$ cm$^{2}$. 
Based on Lee \& Guest \cite{Lee}, the estimated total integrated 
VUV-absorption cross section, 2.3 $\times$ 10$^{-16}$ cm$^{2}$ nm, is larger than the CO ice value, 1.5 $\pm$ 0.1 $\times$ 10$^{-16}$ cm$^{2}$ nm. 
Similar to solid CO, the VUV-absorption cross section of CO gas is very low at 121.6 nm. There are no data available at 157.8 and 
160.8 nm, but the absorption is expected to be very low or zero because the first transition ($\nu_0$) is centered on 154.5 nm.

\begin{table}
\centering
\caption{Transitions observed in the VUV-absorption cross-section spectrum of pure CO ice, deposited at 8 K, in the 115-170 nm range. The peak positions agree with  
previous works, see Mason et al. \cite{Mason}, Lu et al. \cite{Lu}, and  Brith \& Schnepp \cite{Brith}.}
\tiny
\begin{tabular}{ccccc}
\hline
\hline
&\multicolumn{2}{c}{ Band peak}&\multicolumn{2}{c}{ Band area}\\
($\nu$',$\nu$'')&nm&eV&cm$^{2}$nm&cm$^{2}$eV\\
\hline
11,0&127.0&9.76&8.6$\times10^{-20}$&6.6$\times10^{-21}$\\

10,0&128.8&9.62&1.6$\times10^{-19}$&1.2$\times10^{-20}$\\

9,0&131.0&9.46&4.0$\times10^{-19}$&2.9$\times10^{-20}$\\

8,0&132.8&9.33&7.0$\times10^{-19}$&4.9$\times10^{-20}$\\

7,0&135.2&9.17&2.6$\times10^{-18}$&1.8$\times10^{-19}$\\

6,0&137.6&9.01&4.3$\times10^{-18}$&2.8$\times10^{-19}$\\

5,0&140.0&8.85&7.1$\times10^{-18}$&4.5$\times10^{-19}$\\

4,0&142.8&8.68&7.6$\times10^{-18}$&4.6$\times10^{-19}$\\

3,0&146.0&8.49&1.1$\times10^{-17}$&6.5$\times10^{-19}$\\

2,0&149.4&8.29&1.5$\times10^{-17}$&8.4$\times10^{-19}$\\

1,0&153.0&8.10&2.5$\times10^{-17}$&1.3$\times10^{-18}$\\

0,0&156.6&7.91&2.6$\times10^{-17}$&1.3$\times10^{-18}$\\
\hline
\end{tabular}
\\
Error margins corresponding to the values in the second and third column are $\pm$ 0.4 nm and $\pm$ 0.03 eV.
\label{TableCO}
\end{table}

\begin{table}
\centering
\caption{Transitions observed in the VUV-absorption cross-section spectrum of pure CO gas based on data adapted from Lee \& Guest \cite{Lee}.}
\tiny
\begin{tabular}{ccccc}
\hline
\hline
&\multicolumn{2}{c}{ Band peak}&\multicolumn{2}{c}{ Band area}\\
($\nu$',$\nu$'')& nm& eV& cm$^{2}$nm& cm$^{2}$eV\\
\hline
14,0&121.6&10.19&1.5$\times10^{-19}$&1.2$\times10^{-20}$\\

13,0&123.1&10.07&2.3$\times10^{-19}$&1.9$\times10^{-20}$\\

12,0&124.7&9.93&6.6$\times10^{-19}$&5.3$\times10^{-20}$\\

11,0&126.5&9.79&7.5$\times10^{-19}$&5.8$\times10^{-20}$\\

10,0&128.4&9.65&1.3$\times10^{-18}$&9.7$\times10^{-20}$\\

9,0&130.3&9.51&2.4$\times10^{-18}$&1.8$\times10^{-19}$\\

8,0&132.3&9.36&6.5$\times10^{-18}$&4.4$\times10^{-19}$\\

7,0&134.7&9.20&6.8$\times10^{-18}$&4.8$\times10^{-19}$\\

6,0&136.9&9.05&1.2$\times10^{-17}$&8.2$\times10^{-19}$\\

5,0&139.4&8.89&2.1$\times10^{-17}$&1.3$\times10^{-18}$\\

4,0&142.1&8.72&2.6$\times10^{-17}$&1.7$\times10^{-18}$\\

3,0&144.9&8.55&2.7$\times10^{-17}$&1.6$\times10^{-18}$\\

2,0&147.9&8.38&3.3$\times10^{-17}$&1.9$\times10^{-18}$\\

1,0&151.1&8.20&3.0$\times10^{-17}$&1.6$\times10^{-18}$\\

0,0&154.4&8.03&2.3$\times10^{-17}$&1.2$\times10^{-18}$\\
\hline
\end{tabular}
\label{TableCOgas}
\end{table}

\subsection{Solid water}

The ground state and bond energy of H$_{2}$O are \~X$^{1}$A$_{1}$ and E$_{b}$(H--OH) = 5.1 eV (Okabe 1978). 

\begin{figure}[ht!]
\centering
\includegraphics[width=\columnwidth]{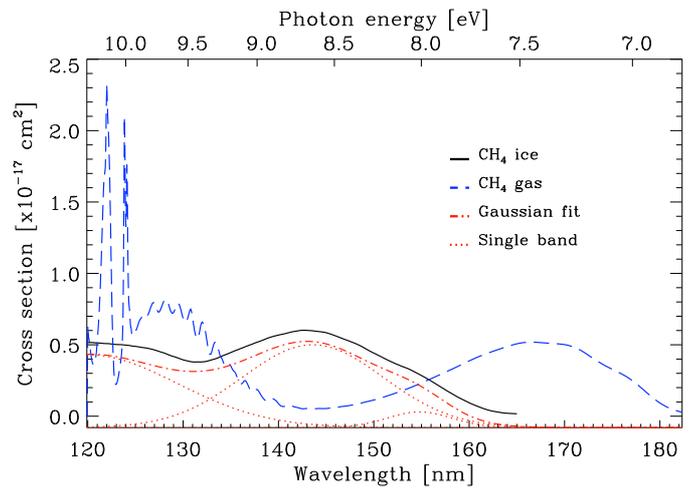}
\caption{VUV-absorption cross section as a function of photon wavelength (bottom X-axis) and VUV-photon energy (top X-axis) of pure H$_{2}$O 
ice deposited at 8 K, black solid trace. The blue dashed trace is the VUV-absorption cross-section spectrum of gas phase H$_{2}$O taken from Mota et al. 
\cite{Mota}. The fit, red dashed-dotted trace, is the sum of three Gaussians, dotted trace. It has been 
vertically offset for clarity.}
\label{H2O}
\end{figure}

The VUV-absorption cross-section spectrum of H$_{2}$O ice is displayed in Fig.~\ref{H2O}. The spectral profile is similar to those 
reported by Lu et al. \cite{Lu2} and Mason et al. \cite{Mason}. The band between 132-163 nm is centered on 142 nm (8.73 eV), 
in agreement with Lu et al. \cite{Lu2} and Mason et al. \cite{Mason}. This band is attributed to the 4a$_{1}$:\~A$^{1}$B$_{1} 
\leftarrow $ 1b$_{1}$:\~X$^{1}$A$_{1}$ transition. The VUV-absorption cross section reaches a value of 6.0$^{+0.4}_{-0.4}$ $\times$ 10$^{-18}$ cm$^{-2}$ 
at this peak, which coincides with the spectrum displayed in Fig. 3 of Mason et al. \cite{Mason}. 
The H$_{2}$O ice spectrum of Lu et al. \cite{Lu2} presents a local minimum at 132 nm, also visible in Fig.~\ref{H2O}. This minimum 
is not observed in the Mason et al. \cite{Mason} data, which is most likely due to the larger scale of their plot. The portion of the band 
in the 120-132 nm range is attributed to the transition \~B$^{1}$A$_{1}$ $\leftarrow$ \~X$^{1}$A$_{1}$, according to Lu et al. \cite{Lu2}. 

The average and the total integrated VUV-absorption cross sections of H$_2$O ice are 3.4$^{+0.2}_{-0.2}$ $\times$ 10$^{-18}$ cm$^{2}$ and 1.8$^{+0.1}_{-0.1}$ $\times$ 
10$^{-16}$ cm$^{2}$ nm (1.2$^{+0.1}_{-0.1}$ $\times$ 10$^{-17}$ cm$^{2}$ 
eV) in the 120-165 nm (10.33-7.51 eV) spectral region. The VUV-absorption cross sections of H$_{2}$O ice at 121.6, 157.8, and 
160.8 nm are 5.2$^{+0.4}_{-0.4}$ $\times$ 10$^{-18}$ cm$^{2}$, 1.7$^{+0.1}_{-0.1}$ $\times$ 10$^{-18}$ cm$^{2}$, and 0.7$^{+0.05}_{-0.05}$ $\times$ 10$^{-18}$ cm$^{2}$.
Gas-phase data from Mota et al. \cite{Mota} were adapted for comparison with our solid-phase data, see Fig.~\ref{H2O}. Gas and ice data 
were previously compared by Mason et al. \cite{Mason}. The VUV-absorption cross section of H$_{2}$O in the gas phase has an average value of 
3.1 $\times$ 10$^{-18}$ cm$^{2}$. H$_{2}$O gas data were integrated in the 120-182 nm range, giving a value 
for the VUV-absorption cross section of 2.3 $\times$ 10$^{-16}$ cm$^{2}$ nm (1.4 $\times$ 10$^{-17}$ cm$^{2}$ eV), which is higher 
than the VUV-absorption cross section of solid H$_{2}$O. The VUV-absorption cross sections of H$_{2}$O gas at 121.6, 157.8, and 
160.8 nm are 13.6 $\times$ 10$^{-18}$ cm$^{2}$, 3.2 $\times$ 10$^{-18}$ cm$^{2}$, and 4.1 $\times$ 10$^{-18}$ cm$^{2}$, 
which is also higher than the solid-phase measurements. 

\subsection{Solid methanol}

The ground state and bond energy of CH$_{3}$OH are X$^{1}$A' and E$_{b}$(H--CH$_{2}$OH) = 4.0 eV (Darwent 1970).

\begin{figure}[ht!]
\centering
\includegraphics[width=\columnwidth]{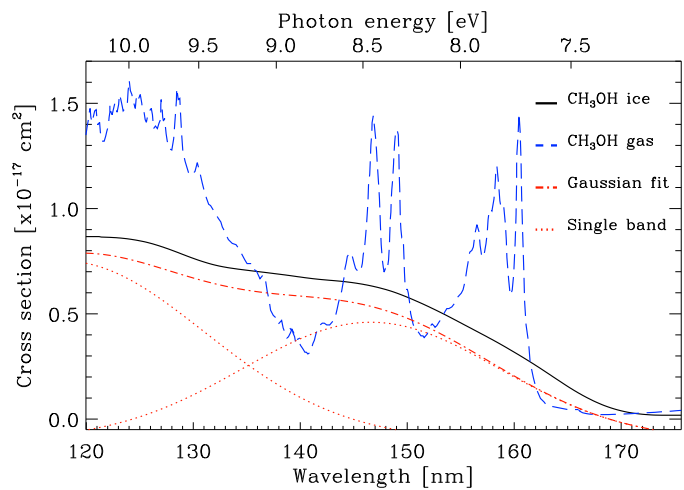}
\caption{VUV-absorption cross section as a function of photon wavelength (bottom X-axis) and VUV-photon energy (top X-axis) of pure CH$_{3}$OH 
ice deposited at 8 K, black solid trace. The blue dashed trace is the VUV-absorption cross-section spectrum of gas phase CH$_{3}$OH adapted from Nee et al. 
\cite{Nee}. The fit, red dashed-dotted trace, is the sum of two Gaussians, dotted trace. It has been vertically offset for clarity. }
\label{CH3OH}
\end{figure}

Fig.~\ref{CH3OH} shows the VUV-absorption cross section of CH$_{3}$OH as a function of wavelength and photon energy. The spectrum 
profile is similar to the one reported by Kuo et al. \cite{Kuo}, a decreasing continuum for longer wavelengths without distinct local maximum. 
These authors found three possible broad bands centered on 147 nm (8.43 eV), 118 nm (10.50 eV), and 106 nm (11.68 eV), the latter 
is beyond our spectral range. We observed the 147 nm peak (associated to the 2$^{1}$A'' $\leftarrow$ 
X$^{1}$A' molecular transition) as well as part of the 118 nm band ( corresponding to the 3$^{1}$A'' $\leftarrow$ X$^{1}$A' molecular 
transition), but due to the decreasing VUV-flux below 120 nm it was not possible to confirm the exact position of this peak. 

The average and the total integrated VUV-absorption cross sections of solid CH$_3$OH are 4.4$^{+0.4}_{-0.7}$ $\times$ 10$^{-18}$ cm$^{2}$ and 
2.7$^{+0.2}_{-0.4}$ $\times$ 10$^{-16}$ cm$^{2}$ nm (1.8$^{+0.1}_{-0.3}$ $\times$ 10$^{-17}$ cm$^{2}$ eV) 
in the 120-173 nm (10.33-7.16 eV) spectral region. The VUV-absorption cross sections of CH$_{3}$OH ice at 121.6 nm, 157.8 and 
160.8 nm are 8.6$^{+0.7}_{-1.3}$ $\times$ 10$^{-18}$ cm$^{2}$, 3.8$^{+0.3}_{-0.6}$ $\times$ 10$^{-18}$ cm$^{2}$, and 2.9$^{+0.2}_{-0.4}$ $\times$ 10$^{-18}$ cm$^{2}$. 
Gas phase data from Nee et al. \cite{Nee} were used for comparison with our solid-phase spectrum, see Fig.~\ref{CH3OH}. Pure gas and solid CH$_3$OH, 
and CH$_{3}$OH in Ar or Kr matrices were compared by Kuo et al. \cite{Kuo}. CH$_{3}$OH gas has a vibrational peak 
profile throughtout the 120-173 nm range, which is absent from in CH$_{3}$OH ice. The VUV-absorption cross section of CH$_{3}$OH in the gas 
phase has an average value of 8.9 $\times$ 10$^{-18}$ cm$^{2}$. CH$_{3}$OH gas data were 
integrated in the 120-173 nm range, giving a value for the VUV-absorption cross section of 3.7 $\times$ 10$^{-16}$ cm$^{2}$ nm 
(2.5 $\times$ 10$^{-17}$ cm$^{2}$ eV), which is higher than the VUV-absorption cross section (2.7$^{+0.2}_{-0.4}$ $\times$ 10$^{-16}$ cm$^{2}$ nm) 
for solid CH$_{3}$OH. The VUV-absorption cross sections of CH$_{3}$OH gas at 121.6 nm, 157.8, and 160.8 nm are 13.2 $\times$ 10$^{-18}$ cm$^{2}$, 9.9 
$\times$ 10$^{-18}$ cm$^{2}$, and 8.6 $\times$ 10$^{-18}$ cm$^{2}$, which are also higher than the ice-phase measurements.

\subsection{Solid ammonia}

The ground state of NH$_{3}$ is \~X$^{1}$A$_{1}$ and its bond energy is E$_{b}$(H--NH$_{2}$) = 4.4 eV (Okabe 1978). 

\begin{figure}[ht!]
\centering
\includegraphics[width=\columnwidth]{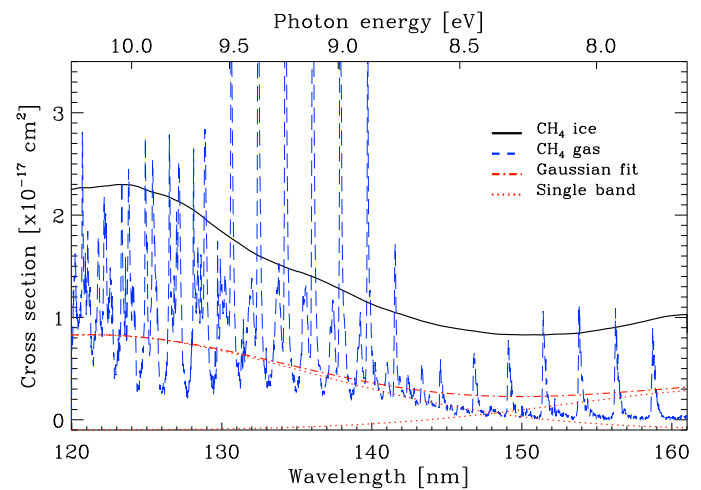}
\caption{VUV-absorption cross section as a function of photon wavelength (bottom X-axis) and VUV-photon energy (top X-axis) of pure NH$_{3}$ 
ice deposited at 8 K, black solid trace. The blue dashed trace is the VUV-absorption cross-section spectrum of gas phase NH$_{3}$ adapted from  Cheng et al. 
\cite{Cheng} and Wu et al. \cite{Wu2}. The fit, red dashed-dotted trace, is the sum of two Gaussians, dotted trace. It has been vertically offset for clarity.}
\label{NH3}
\end{figure}

Fig.~\ref{NH3} displays the VUV-absorption cross section of NH$_{3}$ as a function of the photon wavelength and photon energy. It presents 
a continuum with two broad absorption bands between 120-151 nm (10.33-8.21 eV) and 151-163 nm (8.21-7.60 eV) in this 
wavelength region, without narrow bands associated to vibrational structure. Mason et al. \cite{Mason} found 
a broad band centered on 177 nm (7.00 eV) in the 146-225 nm (5.50-8.50 eV) spectral range. Lu et al. \cite{Lu2} observed 
the same feature centered on 179 nm (6.94 eV). We observed only a portion of this feature because of the low
VUV-flux of the MDHL in the 163-180 nm (7.60-6.88 eV) spectral range. This band is associated to the \~A$^{1}$A$_{2}$'' $\leftarrow$ \~X$^{1}$A$_{1}$ 
molecular transition. The minimum we observed at 151 nm (8.21 eV) is the same as in the above-cited works. At this minimum, the VUV-absorption cross 
section reaches a value of 3.3 $\pm$ 0.2 $\times$ 10$^{-18}$ cm$^{2}$, higher than the 1.9 $\times$ 10$^{-18}$ cm$^{2}$ value reported in Fig. 10 
of Mason et al. \cite{Mason}. We also obtained a higher value at 128.4 nm (9.65 eV), 8.1 $\pm$ 0.5 $\times$ 10$^{-18}$ cm$^{2}$ compared with 
3.8 $\times$ 10$^{-18}$ cm$^{2}$ by Mason et al. \cite{Mason}. This difference is most likely due to the different method employed to estimate 
the ice column density, infrared spectroscopy or laser interferometry. In addition, a broad band centered on 121.2 nm (10.23 eV) was 
observed, in agreement with Lu et al. \cite{Lu2}, probably associated to the \~B $\leftarrow$ \~X molecular transition. 

The average and the total integrated VUV-absorption cross sections of solid NH$_3$ are 4.0$^{+0.3}_{-0.5}$ $\times$ 10$^{-18}$ cm$^{2}$ and 2.2$^{+0.2}_{-0.3}$ 
$\times$ 10$^{-16}$ cm$^{2}$ nm (1.5$^{+0.1}_{-0.2}$ $\times$ 10$^{-17}$ cm$^{2}$ eV) 
in the 120-161 nm (10.33-7.70 eV) spectral region. The VUV-absorption cross sections of NH$_{3}$ ice at 121.6 nm, 157.8 and 
160.8 nm are 9.1$^{+0.7}_{-1.1}$ $\times$ 10$^{-18}$ cm$^{2}$, 3.8$^{+0.3}_{-0.5}$ $\times$ 10$^{-18}$ cm$^{2}$, and 4.1$^{+0.3}_{-0.5}$ $\times$ 10$^{-18}$ cm$^{2}$. 
Gas-phase data from Cheng et al. \cite{Cheng} and Wu et al. \cite{Wu2} were adapted, 
see Fig.~\ref{CH3OH}. At least qualitatively, our result is compatible with Mason et al. \cite{Mason}. The VUV-absorption cross section of 
NH$_{3}$ in the gas phase has an average value of 6.1 $\times$ 10$^{-18}$ cm$^{2}$. NH$_{3}$ gas data were 
integrated in the 120-161 nm range, giving a value of 2.5 $\times$ 10$^{-16}$ cm$^{2}$ nm (1.8$\times$ 10$^{-17}$ cm$^{2}$ eV), slightly higher than 
the VUV-absorption cross section of solid NH$_{3}$ (2.2$^{+0.2}_{-0.3}$ $\times$ 10$^{-16}$ cm$^{2}$ nm). The VUV-absorption cross 
sections of NH$_{3}$ gas at 121.6, 157.8 and 160.8 nm are 9.8 $\times$ 10$^{-18}$ cm$^{2}$, 0.1 $\times$ 
10$^{-18}$ cm$^{2}$, and 0.3 $\times$ 10$^{-18}$ cm$^{2}$; these values are lower than the ice-phase measurements, except for the Ly-$\alpha$ photon wavelength (121.6 nm).

\subsection{Solid hydrogen sulfide}

The ground state and bond energy of H$_{2}$S are \~X$^{1}$A$_{1}$ and E$_{b}$(H--SH) = 3.91 eV (Okabe 1978). 

\begin{figure}[ht!]
\centering
\includegraphics[width=\columnwidth]{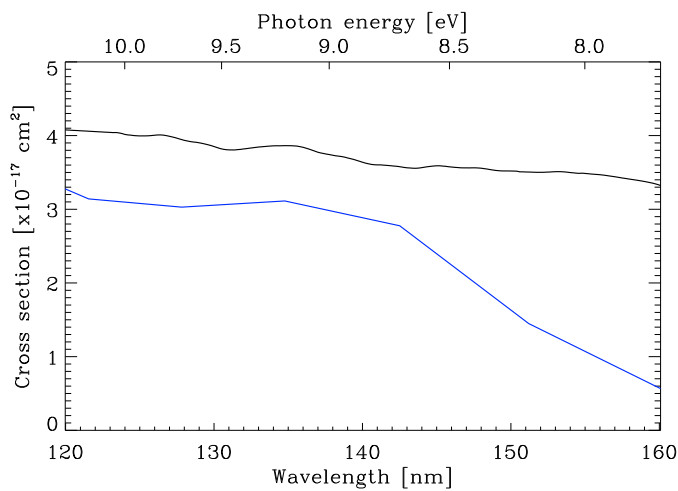}
\caption{VUV-absorption cross section as a function of wavelength (bottom X-axis) and photon energy (top X-axis) of pure H$_{2}$S ice deposited at 8 K, 
black trace. The blue trace is the VUV-absorption cross-section spectrum of gas phase H$_{2}$S adapted from Feng et al. \cite{Feng}}
\label{H2S}
\end{figure}

Fig.~\ref{H2S} shows the VUV-absorption cross section of H$_{2}$S as a function of the wavelength and photon energy. Owing to the high 
VUV-absorption of solid H$_{2}$S, very thin ice samples were deposited to obtain a proper VUV-spectrum. 
H$_{2}$S gas has a nearly constant VUV-absorption cross section in the 120-150 nm (10.33-8.26 eV) range, between 150-180 nm 
(8.26-6.88 eV) it decreases, and in the 180-250 nm (6.88-4.95 eV) region it presents a maximum at 187 nm (6.63 eV), see Okabe \cite{Okabe}. The VUV-absorption cross section of H$_{2}$S 
is almost constant in the 120-143 nm (10.33-8.67 eV) range and 
decreases in the 143-173 nm (8.67-7.16 eV) range, with a maximum at 200 nm (6.19 eV) according to
Feng et al. \cite{Feng}. The $^{1}$B$_{1}$ 
$\leftarrow$ $^{1}$A$_{1}$ molecular transition was identified at 139.1 nm (8.91 eV) by Price \& Simpson \cite{Price} and confirmed by Gallo \& Innes \cite{Gallo}.

The total integrated VUV-absorption cross section of H$_2$S ice has a value of 1.5$^{+0.1}_{-0.3}$ $\times$ 10$^{-15}$ cm$^{2}$ nm (4.3$^{+0.3}_{-1.0}$ $\times$ 10$^{-16}$ cm$^{2}$ 
eV) in the 120-160 nm (10.33-7.74 eV) spectral region. The VUV-absorption cross sections of the same ice at  121.6, 157.8, and 
160.8 nm are 4.2$^{+0.3}_{-1.0}$ $\times$ 10$^{-17}$ cm$^{2}$, 3.4$^{+0.3}_{-0.8}$ $\times$ 10$^{-17}$ cm$^{2}$, and 3.3$^{+0.3}_{-0.8}$ $\times$ 10$^{-17}$ cm$^{2}$.
Gas-phase data from Feng et al. \cite{Feng} were adapted, see Fig.~\ref{H2S}. The spectrum 
of the solid is almost constant, only a slight decrease is observed as the wavelength increases. The gas-phase spectrum presents 
an abrupt decrease in the cross section starting near 140 nm, according to Feng et al. \cite{Feng}. In contrast to the Okabe \cite{Okabe} 
data, this spectrum of the gas shows no vibrational structure, probably because of its low resolution. The average VUV-absorption 
cross section of H$_{2}$S gas reported by Feng et al. \cite{Feng}, 3.1 $\times$ 10$^{-17}$ cm$^{2}$, disagrees by one order of magnitude 
with the 3-4 $\times$ 10$^{-18}$ cm$^{2}$ value of Okabe \cite{Okabe}. We estimated a value of 3.9$^{+0.3}_{-0.9}$ $\times$ 10$^{-17}$ cm$^{2}$ for the solid in 
the same range, which is close to the Feng et al. \cite{Feng} gas-phase value.

H$_{2}$S gas data were integrated 
in the 120-160 nm range, giving a value of 10.2 $\times$ 10$^{-16}$ cm$^{2}$ nm (8.0 $\times$ 10$^{-17}$ cm$^{2}$ eV), which is lower 
than the VUV-absorption cross section of solid H$_{2}$S. The VUV-absorption cross sections of H$_{2}$S gas at 121.6, 157.8, and 
160.8 nm are 31.0 $\times$ 10$^{-18}$ cm$^{2}$, 7.9 $\times$ 10$^{-18}$ cm$^{2}$, and 4.9 $\times$ 10$^{-18}$ cm$^{2}$, 
which are also lower than the solid-phase values reported above.

\subsection{Comparison between all the ice species}

\begin{figure}[ht!]
\centering
\includegraphics[width=\columnwidth]{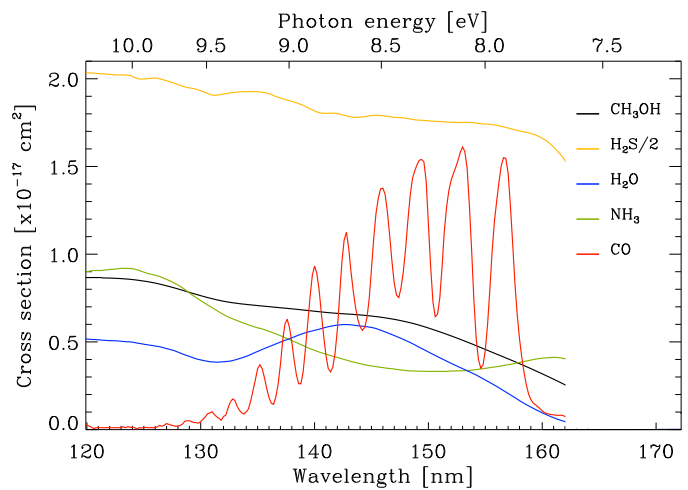}
\caption{VUV-absorption cross section as a function of wavelength (bottom x-axis) and photon energy (top x-axis) of all the 
pure ice species studied in this work, deposited at 8 K.}
\label{todas}
\end{figure}

Fig.~\ref{todas} shows a comparison of the VUV-absorption cross section for all the ice species deposited at 8 K, represented in the 
same linear scale. The most absorbing molecule is H$_{2}$S. H$_{2}$O has the lowest average absorption in this 
range. All the represented species absorb VUV-light significantly in the 120-163 nm (10.33-7.60 eV) range, only the CO absorption is negligible at 
the Ly-$\alpha$ wavelength. Except for CO, the absorption of the reported ice species in the H$_{2}$ molecular emission region between 157-161 nm is lower than 
the Ly-$\alpha$ absorption. Table ~\ref{tableInt} summarizes the comparison between the VUV-absorption cross sections of all the species in 
the gas and solid phase. 

\begin{table*}
\centering
\caption{Comparison between the VUV-absorption cross sections in the 120-160 nm range of all the pure species in the gas and solid (deposited at 8 K) phase. 
Total Int. is the total integrated VUV-absorption cross section in this range, Avg. is the average VUV-absorption cross section. 
Ly-$\alpha$ and LBS are the average VUV-absorption cross sections in the 120.8-122.6 nm and 132-162 nm range normalized by the  
photon flux. The last three columns are the VUV-absorption cross sections at wavelengths 121.6 nm (corresponding to the maximum intensity 
of the Ly-$\alpha$ peak), 157.8 and 160.8 nm (main peaks of the Lyman band system). }
\small
\begin{tabular}{ccccccccc}
\hline
\hline
& Species& Total Int.& Avg.& Ly-$\alpha$& LBS& 121.6 nm& 157.8 nm& 160.8 nm\\
&& [$\times$ 10$^{-16}$ cm$^{2}$ nm]&\multicolumn{6}{c}{ [$\times$ 10$^{-18}$ cm$^{2}$]}\\
\hline
\multirow{10}{*}{\rotatebox{90}{ Solid phase}}&&&&&&&&\\
&CO&1.5$^{+0.1}_{-0.1}$ &4.7$^{+0.4}_{-0.4}$&0.1$^{+0.01}_{-0.01}$&11$^{+0.9}_{-0.9}$&0.1$^{+0.01}_{-0.01}$&6.6$^{+0.5}_{-0.5}$&0.9$^{+0.07}_{-0.07}$\\
&&&&&&&&\\
&H$_{2}$O&1.8$^{+0.1}_{-0.1}$&3.4$^{+0.2}_{-0.2}$&5.1$^{+0.4}_{-0.4}$&4.0$^{+0.3}_{-0.3}$&5.2$^{+0.4}_{-0.4}$&1.7$^{+0.1}_{-0.1}$&0.7$^{+0.05}_{-0.05}$\\
&&&&&&&&\\
&CH$_{3}$OH&2.2$^{+0.2}_{-0.3}$&4.4$^{+0.4}_{-0.7}$&6.7$^{+0.5}_{-1.0}$&6.4$^{+0.5}_{-1.0}$&8.6$^{+0.7}_{-1.3}$&3.8$^{+0.3}_{-0.6}$&2.9$^{+0.2}_{-0.4}$\\
&&&&&&&&\\
&NH$_{3}$&2.2$^{+0.1}_{-0.3}$&4.0$^{+0.3}_{-0.5}$&9.1$^{+0.7}_{-1.1}$&3.6$^{+0.3}_{-0.5}$&9.1$^{+0.7}_{-1.1}$&3.8$^{+0.3}_{-0.5}$&4.1$^{+0.3}_{-0.5}$\\
&&&&&&&&\\
&H$_{2}$S&15$^{+1}_{-3}$&39$^{+3}_{-9}$&41$^{+3}_{-10}$&34.0$^{+3}_{-8}$&42$^{+3}_{-10}$&34$^{+3}_{-8}$&33$^{+3}_{-8}$\\
\hline
\hline
\multirow{10}{*}{\rotatebox{90}{ Gas phase}}&&&&&&&&\\
&CO&2.3&20.7&--&--&--&--&--\\
&&&&&&&&\\
&H$_{2}$O&2.3&3.1&--&--&13.6&3.2&4.1\\
&&&&&&&&\\
&CH$_{3}$OH&3.4&8.2&--&--&13.6&8.5&10.0\\
&&&&&&&&\\
&NH$_{3}$&2.5&6.1&--&--&9.8&0.1&0.3\\
&&&&&&&&\\
&H$_{2}$S&10.2&31.0&--&--&31.0&7.9&4.9\\
\hline
\end{tabular}
\label{tableInt}
\end{table*}

\section{Astrophysical implications}
\label{Astro}

Far-ultraviolet observations of IC 63, an emission/reflection nebula illuminated by the B0.5 IV star $\gamma$Cas, with the \emph{Hopkins 
Ultraviolet Telescope} (HUT) on the central nebular position, have revealed a VUV-emission spectrum very similar 
to the spectrum of our VUV-lamp, see Fig. 4 of France et al. \cite{France}, which means that the VUV-spectrum that interacts 
with the ice sample mimicks the molecular hydrogen photoexcitation observed toward this photodissociation region (PDR) in the local interstellar medium. 
This spectrum is similar to that emitted by the MDHL with a H$_{2}$ pressure of 0.2 mbar. 

\begin{figure}[ht!]
\centering
\includegraphics[width=\columnwidth]{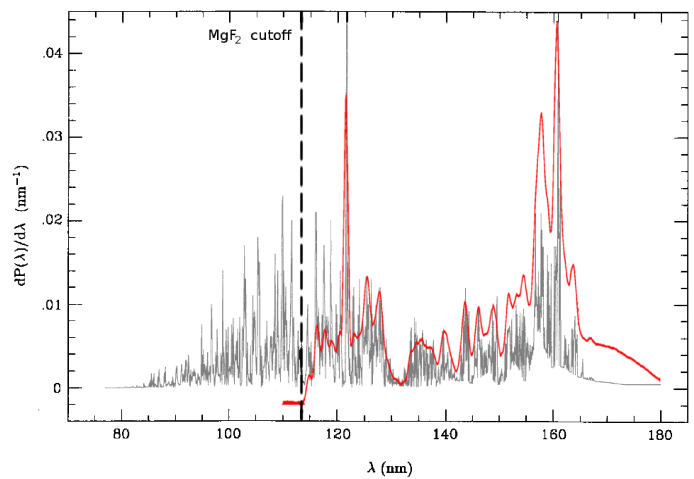}
\caption{Calculated spectrum of ultraviolet photons created in the interior of dense molecular clouds by impact excitation of molecular hydrogen by cosmic ray ionization, 
adapted from Gredel et al. \cite{Gredel}, in the background. VUV-emission spectrum of the MDHL used in this work, red foreground trace.}
\label{2UV}
\end{figure}

Ice mantle build-up takes place in cold environments such as dark cloud interiors, where the radiation acting on dust grains is the secondary UV field produced by cosmic ray 
ionization of H$_2$. These are the conditions mimicked in our experiments: low density and low temperature, and UV photons impinging on ice mantles. The secondary UV field 
calculated by Gredel et al. \cite{Gredel} is very similar to the emission spectrum of our lamp, see Fig.~\ref{2UV}, except for photons with wavelengths shorter than about 114 nm, 
which are not transmitted by the MgF$_2$ window used as interface between the MDHL and the ISAC set-up. 

The reported VUV-absorption cross sections allow a more quantitative study of photon absorption in ice mantles. Based on our data, the VUV-light reaching an interstellar ice 
mantle must have an energy higher than 7.00, 7.60, and 7.08 eV, to be absorbed by solid CO, H$_{2}$O, and CH$_{3}$OH, respectively. The full absorption spectrum 
in the VUV measured by other authors provides an absorption threshold above 6.20 and 5.16 eV for NH$_{3}$ and H$_{2}$S, respectively, 
see Mason et al. \cite{Mason} and Feng et al. \cite{Feng}.\\

The ice penetration depth of photons with a given wavelength, or the equivalent absorbing ice column  density of a species in the solid 
phase, can be calculated from the VUV-absorption cross section following Eq.~\ref{2}. Table ~\ref{penetration} summarizes the penetration depth of the 
ice species for an absorbed photon flux of 95\% and 99\% using the cross section value at Ly-$\alpha$, the average value of the cross section in the 120-160 nm range, and 
the maximum value of the cross section in the same range. 

\begin{table}[ht!]
\centering
\caption{Penetration depth, expressed as absorbing column density, of the different pure ice species deposited at 8 K, corresponding to an absorbed 
photon flux of 95\% and 99\%. Ly-$\alpha$ corresponds to the cross section at the Ly-$\alpha$ wavelength, 121.6 nm; in the case 
of CO, the upper limit in the cross section leads to a lower limit in the absorbing column density. Avg. corresponds to the average 
value of the cross section in the 120-160 nm range. Max. corresponds to the maximum value of the cross section in the same wavelength range. Errors in the penetration 
depth correspond to those estimated for the VUV-absorption cross sections of the different ices.}
\tiny
\begin{tabular}{ccccccc}
\hline
\hline
&\multicolumn{3}{c}{ 95\% photon absorption}&\multicolumn{3}{c}{ 99\% photon absorption}\\
Species& Ly-$\alpha$& Avg.& Max.& Ly-$\alpha$& Avg.& Max.\\
&\multicolumn{3}{c}{ ($\times$10$^{17}$ molecule cm$^{-2}$)}&\multicolumn{3}{c}{($\times$10$^{17}$ molecule cm$^{-2}$)}\\
\hline
CO&$\geq$150&6.4&1.9&$\geq$230&9.8&2.9\\
H$_{2}$O&5.8&8.3&4.9&8.9&13.0&7.7\\
CH$_3$OH&3.5&5.7&3.4&5.4&8.7&5.3\\
NH$_3$&3.3&5.5&3.2&5.1&8.5&5.0\\
H$_2$S&0.73&0.81&0.73&1.1&1.2&1.1\\
\hline
\end{tabular}
\label{penetration}
\end{table}

The VUV-absorption of ice mantles is low compared with that of the dust grain cores, and therefore difficult to observe directly, but it 
is an essential parameter quantitatively for studing the photoprocessing and photodesorption of molecular species in the ice mantle. The desorbing photoproducts can 
be detected in observations of the gas phase. The reported measurements of the cross sections are thus needed to estimate the absorption of icy grain mantles in the photon 
range where molecules 
are photodissociated or photodesorbed, which often leads to the formation of photoproducts. These results can be used as input in 
models that predict the processing of ice mantles that are exposed to an interstellar VUV-field. They can be used, for instance, to predict the gas phase abundances of 
molecules photodesorbed from the ice mantles, as we discuss below.

There is a clear correspondence between the photodesorption rates of CO ice measured at different photon energies (Fayolle et al. 2011) and the VUV-absorption spectrum of CO ice 
(Lu et al. 2009, this work) for the same photon energies. 
This indicates that photodesorption of some ice species like CO and N$_2$ is mainly driven by a desorption induced by electronic transition (DIET) process 
(Fayolle et al. 2011; 2013). The lowest photodesorption reported by Fayolle et al. \cite{Fayolle1} at the Ly-$\alpha$ wavelength (121.6 nm) is \textless 6 $\times$ 
10$^{-3}$ molecules per incident photon, coinciding with the low VUV-absorption at this wavelength in Fig.~\ref{CO}, and the maximum in the photodesorption occurs 
approximately at $\sim$ 151.2 nm, near the most intense VUV-absorption band, see Fig.~\ref{CO}. The photodesorption rate per absorbed photon in the [$\lambda_{i}$,$\lambda_{f}$] 
wavelength range, $R^{\rm abs}_{\rm ph-des}$, can differ significantly from the photodesorption rate per incident photon, $R^{\rm inc}_{\rm ph-des}$. It can be estimated as follows:
\begin{eqnarray}
\centering
 R^{\rm abs}_{\rm ph-des} &=& \frac{\Delta N}{I_{abs}} \quad and \quad  R^{\rm inc}_{\rm ph-des} = \frac{\Delta N}{I_0} \nonumber \\
\nonumber \\
\nonumber \\
 R^{\rm abs}_{\rm ph-des} &=& \frac{I_0}{I_{abs}} \; R^{\rm inc}_{\rm ph-des},
\label{RR}
\end{eqnarray}
where
\begin{eqnarray}
I_{abs} &=& \displaystyle\sum\limits_{\lambda_i}^{\lambda_f} \quad I_0(\lambda) - I(\lambda) = \displaystyle\sum\limits_{\lambda_i}^{\lambda_f} \quad I_0(\lambda)(1 - e^{- \sigma(\lambda) N}), \nonumber
\end{eqnarray}
and $\Delta N$ is the column density decrease for a given irradiation time in molecules cm$^{-2}$ s$^{-1}$, $I_0$ is the total photon flux emitted (Fayolle et al. 2011 
reports $\sim$ 1.7 $\times$ 10$^{14}$ photons cm$^{-2}$ s$^{-1}$, in our experiments $\sim$ 2.0 $\times$ 10$^{14}$ photons cm$^{-2}$ s$^{-1}$), $I_{abs}$ is the total photon flux 
absorbed by the ice, $I_0(\lambda)$ is the photon flux emitted at wavelength $\lambda$, $\sigma(\lambda)$ is the VUV absorption cross section at the same wavelength, and $N$ is the 
column density of the ice sample. 

Fayolle et al. \cite{Fayolle1} reported $R^{\rm inc}_{\rm ph-des}$ = 5 $\times$ 10$^{-2}$ molecules per incident photon for monochromatic $\sim$ 8.2 eV light irradiation of a 9-10 ML 
ice column density of CO. 
We also estimated the value of $R^{\rm inc.}_{\rm ph-des}$ in our CO irradiation experiment using the MDHL continuum emission source with an average photon energy of 8.6 eV; this 
gave a value of 5.1 $\pm$ 0.2 $\times$ 10$^{-2}$ for an ice column density of 223 ML, in agreement with Mu\~noz Caro et al. \cite{Caro1}. Therefore, similar values of 
$R^{\rm inc}_{\rm ph-des}$ are obtained using either, monochromatic or continuum emission sources, provided that the photon energy of the former is near the average photon energy of 
the latter, this is discussed below. 
Using Eq.~\ref{RR}, we estimated the $R^{\rm abs}_{\rm ph-des}$ values in both experiments; they are summarized in Table \ref{Rabs} for a column density of 5 ML (only the photons 
absorbed in the top 5 $\pm$ 1 ML participate in the photodesorption) for three monochromatic photon energies in the Fayolle et al. \cite{Fayolle1} work, selected to coincide with 
Lyman-$\alpha$ (10.2 eV), the maximum cross section (9.2 eV energy), and the monochromatic energy of 8.2 eV, that is, the one closer to the average photon energy of the MDHL at 8.6 eV.  

\begin{table}[ht!]
\centering
\caption{VUV-absorption cross sections 
for different irradiation energies. R$_{inc}$ values correspond to Fayolle et al. (2011), except for 8.6 eV, which is the average photon energy of our MDHL and the corresponding value 
of the cross section is the average VUV-absorption cross section of pure CO ice, deposited at 8 K, in the 120-160 nm range. $R^{\rm abs}_{\rm ph-des}$ values for a column density of 
about 5 ML, i.e. where the absorbed photons contribute to a photodesorption event (Mu\~noz Caro et al. 2010; Fayolle et al. 2011).}
\tiny
\begin{tabular}{cccc}
\hline
Irrad. energy& $\sigma$& R$^{\rm inc}_{\rm ph-des}$& R$^{\rm abs}_{\rm ph-des}$\\
eV&cm$^{2}$&molec./photon$_{inc}$&  molec./photon$_{abs}$\\
\hline
10.2 & 1.1 $\times$ 10$^{-19}$& 6.9 $\pm$ 2.4 $\times$ 10$^{-3}$& 12.5 $\pm$ 4.4\\
9.2 &  2.8 $\times$ 10$^{-18}$& 1.3 $\pm$ 0.91 $\times$ 10$^{-2}$& 0.9 $\pm$ 0.6\\
\hspace{-0.4cm}
\textdagger $\:$ 8.2 &  9.3 $\times$ 10$^{-18}$& 5 $\times$ 10$^{-2}$& 1.1 \\
\hline
 8.6 &  4.7 $\times$ 10$^{-18}$& 5.1 $\pm$ 0.2 $\times$ 10$^{-2}$& 2.5 $\pm$ 0.1\\
\hline
\end{tabular}
\\
\textdagger Peak yield value at $\sim$ 8.2 eV, see Fayolle et al. (2011).\\
\label{Rabs}
\end{table}

Irradiation of CO ice using our VUV-lamp, with a broad-band energy distribution and mean energy of $\sim$ 8.6 eV, gives similar $R^{\rm inc}_{\rm ph-des}$ values, but different 
$R^{\rm abs}_{\rm ph-des}$ values than irradiation with a $\sim$ 8.2 eV energy monochromatic light source (Fayolle et al. 2011). This suggests that in this particular case, the 
photodesorption rate per incident photon does not depend much on the photon energy distribution, but this coincidence is only by chance, since the corresponding 
$R^{\rm abs}_{\rm ph-des}$ values differ significantly.

The properties of individual molecular componentes of interstellar ice analogs have been studied over the years (e.g., Sandford et al. 1988; Sandford \& Allamandola 1990; Gerakines 
et al. 1996; Escribano et al. 2013). Our VUV-absorption spectra can be directly applied to astrophysical icy environments practically made 
of a single compound, for example, either CO$_2$ or H$_2$O largely dominate the composition of different regions at the south pole of Mars (Bibring et al. 2004). Nevertheless, interstellar 
ice mantles are either thought to be a mixture of different species or to present a layered structure (e.g., Gerakines et al. 1995, Dartois et al. 1999, Pontoppidan et al. 2008, 
Boogert et al. 2011, \"Oberg et al. 2011, Kim et al. 2012). Our study on pure ices can be used to estimate the absorption of multilayered ice mantles, but {\it a priori}, it cannot 
be extrapolated to ice mixtures. A follow-up of this work will include ice mixtures and 
VUV-spectroscopy at ice temperatures above 8 K. 

\section{Conclusions}
\label{Conclusions}

Several conclusions can be drawn from our experimental work; they are summarized as follows:

\begin{itemize}
\item The combination of infrared (FTIR) spectroscopy in transmission for measuring the deposited ice column density and VUV-spectroscopy 
allowed us to determine more accurate VUV-absorption cross-section values of interstellar ice analogs with an error within 16 \%, 15 \%, 23 \%, 19 \%, and 32 \% for CO, H$_2$O, 
CH$_3$OH, NH$_3$, and H$_2$S. The errors are mainly caused by the ice column density decrease due to VUV-irradiation during VUV spectral acquisition.
\end{itemize}
\begin{itemize}
\item For the first time, the VUV-absorption cross sections of CO, CH$_{3}$OH and H$_{2}$S were measured for the solid phase, with average 
VUV-absorption cross sections of 4.7$^{+0.4}_{-0.4}$ $\times$ 10$^{-18}$ cm$^{2}$, 4.4$^{+0.4}_{-0.7}$ $\times$ 10$^{-18}$ cm$^{2}$, and 39$^{+3}_{-9}$ 
$\times$ 10$^{-18}$ cm$^{2}$. 
The total integrated VUV-absorption cross sections are 1.5$^{+0.1}_{-0.1}$ $\times$ 10$^{-16}$ cm$^{2}$ nm,  2.7$^{+0.2}_{-0.4}$ $\times$ 10$^{-16}$ cm$^{2}$ nm, 
and 1.5$^{+0.1}_{-0.3}$ $\times$ 10$^{-15}$ cm$^{2}$ nm. Our estimated values of the average VUV-absorption cross sections of H$_{2}$O and NH$_{3}$ ices 
3.4$^{+0.2}_{-0.2}$ $\times$ 10$^{-18}$ cm$^{2}$ and 4.0$^{+0.3}_{-0.5}$ $\times$ 10$^{-18}$ cm$^{2}$, are comparable with those reported by Mason et al. \cite{Mason}, which were 
measured using a synchrotron as the emission source.
\end{itemize}
\begin{itemize}
\item The ice samples made of molecules such as H$_{2}$O, CH$_{3}$OH, and NH$_{3}$ present broad absorption bands and similar average
VUV-absorption cross sections between 3.4$^{+0.2}_{-0.2}$ $\times$ 10$^{-18}$ cm$^{2}$ and 4.4$^{+0.4}_{-0.7}$ $\times$ 10$^{-18}$ cm$^{2}$, see Fig. ~\ref{todas}. But H$_{2}$S ice, 
which also displays a continuum spectrum, presents about four times more absorption than the other molecules in the same VUV range. 
\end{itemize}
\begin{itemize}
\item Solid CO displays discrete VUV-absorption bands and very low absorption at the Ly-$\alpha$ wavelength. But the other solid samples present 
high absorption at the Ly-$\alpha$ wavelength.
\end{itemize}
\begin{itemize}
\item For the H$_{2}$O, CH$_{3}$OH, and NH$_{3}$ species, the VUV-absorption range and the total integrated VUV-absorption cross 
section in the gas phase is larger than the solid phase. An exception is H$_{2}$S. 
\end{itemize}
\begin{itemize}
\item The main emission peaks of the MDHL occur at 121.6 nm (Lyman-$\alpha$), 157.8 nm 
and 160.8 nm (molecular H$_2$ bands), see Fig.~\ref{Uno}. It is still a common mistake in the molecular astrophysics community to consider the MDHL as a pure Ly-$\alpha$ photon 
source. 
As we mentioned, the lamp VUV-spectrum affects the photochemistry but not the VUV-spectroscopy in a direct way, since the VUV-emission spectrum was subtracted to measure 
the ice absorption in the same energy range.
\end{itemize}
\begin{itemize}
\item Monitoring of the photon energy distribution and the stability of the irradiation source is important for studing ice photoprocesses.
\end{itemize}
\begin{itemize}
\item The results are satisfactory and demonstrate the viability of the MDHL, which commonly used in irradiation experiments, as a source for VUV spectroscopy of solid samples. 
This method to perform VUV-spectroscopy does not require a synchrotron facility and can be used routinely in the laboratory.
\end{itemize}
\begin{itemize}
\item Our estimates of the VUV-absorption cross sections of polar ice molecules can be used as input in models that simulate the photoprocessing of ice mantles present in cold 
environments, such as dense cloud interiors and circumstellar regions. The data reported in this paper can be applied to estimate the absorption of layered ice mantles, but not to 
ices built up with different species that are intimately mixed.
\end{itemize}

\begin{acknowledgements}
This research was financed by the Spanish MICINN under projects AYA2011-29375 and CONSOLIDER grant CSD2009-00038. This work was partially supported by NSC 
grants NSC99-2112-M-008-011-MY3 and NSC99-2923-M-008-011-MY3, and the NSF Planetary Astronomy Program under Grant AST-1108898.
\end{acknowledgements}


\begin{thebibliography}{99}
\bibitem[(2004)]{Bibring} Bibring, J. -P., Langevin, Y., Poulet, F., et al. 2004, Nature, 428, 627
\bibitem[(1978)]{Boursey} Boursey, E., Chandrasekharan, V., G\"urtler, P., et al. 1978, Phys. Rev. Lett., 41, 1516
\bibitem[(2011)]{Boogert} Boogert, A. C. A., Huard, T. L., Cook, A. M., et al. 2011, ApJ, 729, 92
\bibitem[(1965)]{Brith} Brith, M., \& Schnepp, O. 1965, Molecular Phys., 9, 473
\bibitem[(1992)]{Cecchi92}  Cecchi-Pestellini, \& Aiello, S. 1992, MNRAS, 258, 125
\bibitem[(2010)]{Asper1} Chen, Y.-J., Chu, C.-C, Lin, Y.-C, et al. 2010, Advances in Geosciences, 25, 259
\bibitem[(2013)]{Asper2} Chen, Y.-J., Chuang, K.-Y., \& Mu\~noz Caro, G. M. 2013, ApJ, in press.
\bibitem[(2006)]{Cheng} Cheng, B.-M., Lu, H.-C., Chen, H.-K., et al. 2006, ApJ, 647, 1535
\bibitem[(2011)]{Cheng2} Cheng, B.-M., Chen, H.-F., Lu, H.-C., et al. 2011, ApJSS, 196:3, 6
\bibitem[(2004)]{Collings} Collings, M.P., Anderson, M.A., Chen, R., et al. 2004, Mon. Not. R. Astron. Soc., 354, 1133
\bibitem[(1970)]{Dalgarno} Dalgarno, A., \& Stephens, T. L. 1970, ApJ, 160, L107
\bibitem[(1970)]{Darwent} Darwent, B. deB. 1970, Bond Dissociation Energies in Simple Molecules, National Standard Reference Data 
System, 31, 70-602101
\bibitem[(1999)]{Dartois} Dartois, E., Demyk, K., d'Hendecourt, L., \& Ehrenfreund, P. 1999, A\&A, 351, 1066
\bibitem[(2005)]{DartoisCO2} Dartois, E., Pontoppidan, K., Thi, W.-F., \& Mu\~noz Caro, G. M. 2005, A\&A, 444, L57
\bibitem[(2009)]{Dartois6} Dartois, E. 2009. ASP Conference Series, 414, 411
\bibitem[(2007)]{Dawes} Dawes, A., Mukerji, R. J., Davis, M. P., et al. 2007, J. Chem. Phys., 126, 244711
\bibitem[(1985)]{dHendecourt1} d'Hendecourt, L. B., Allamandola, L. J., \& Greenberg, J. M. 1985. A\&A, 152, 130
\bibitem[(1986)]{dHendecourt2} d'Hendecourt, L. B., Allamandola, L. J., Grim, R. J. A., \& Greenberg, J. M. 1986. A\&A, 158, 119
\bibitem[(1998)]{Ehrenfreund} Ehrenfreund, P., \& van Dishoeck, E. F. 1998, Advances in Space Research, 21, 15 
\bibitem[(2013)]{Escribano} Escribano, R., Mu\~noz Caro, G. M., Cruz-Diaz, G. A., Rodr\'\i{}guez-Lazcano, Y., \& Mat\'e, B. 2013, PNAS, 110, 32, 12899
\bibitem[(2011)]{Fayolle1} Fayolle, E. C., Bertin, M., Romanzin, C., et al. 2011, ApJ Letters, 739, L36
\bibitem[(2013)]{Fayolle2} Fayolle, E. C., Bertin, M., Romanzin, C., et al. 2013, A\&A, 556, A122
\bibitem[(1999)]{Feng} Feng, R., Cooper, G., \& Brion, C. E. 1999, Chem. Phys., 244, 127
\bibitem[(2012)]{Fillion} Fillion, J.-H., Bertin, M., Lekic, A., et al. 2012, EAS Publications Series, 58, 307
\bibitem[(2005)]{France} France, K., Andersson, B.-G., McCandliss, S. R., \& Feldman, P. D. 2005, ApJ, 628, 750
\bibitem[(1975)]{Gallo} Gallo, A. R., \& Innes, K.K. 1975, J. Mol. Spectrosc., 54, 472
\bibitem[(1995)]{Gerakines2} Gerakines, P. A., Schutte, W. A., Greenberg, J. M. \& van Dishoeck, E. F. 1995. A\&A, 296, 810
\bibitem[(1996)]{Gerakines} Gerakines, P. A., Schutte, W. A. \& Ehrenfreund, P., 1996, A\&A, 312, 289
\bibitem[(2004)]{Gibb} Gibb, E. L., Whittet, D. C. B., Boogert, A. C. A., \& Tielens, A. G. G. M. 2004, ApJ, 151, 35
\bibitem[(1989)]{Gredel} Gredel, R., Lepp, S., \& Dalgarno, A. 1989, ApJ, 347, 289
\bibitem[(1968)]{Hudson} Hudson, R. D., \& Carter, V. L. 1968, J. Opt. Soc. Am., 58, 227
\bibitem[(1954)]{Inn} Inn, E. C. Y. 1954, Spectrochimica Acta, 7, 65
\bibitem[(1974)]{Jiang} Jiang, G. J., Person, W. B., \& Brown, K. G. 1975, J. Chem. Phys., 75, 4198
\bibitem[(2011)]{Antonio} Jim\'enez-Escobar, A., \& Mu\~noz Caro, G. M. 2011, A\&A, 536, 11
\bibitem[(2012)]{Kim} Kim, H. J., Evans II, N. J., Dunham, M. M., Lee, J. -E., \& Pontoppidan, K. M. 2012, ApJ, 758, 38
\bibitem[(2005)]{Knez} Knez, C., Boogert, A. C. A., Pontoppidan, K. M., et al. 2005, ApJ, 635, L145
\bibitem[(2007)]{Kuo} Kuo, Y.-P., Lu, H.-C., Wu, Y.-J., Cheng, B.-M., \& Ogilvie, J. F. 2007, Chem. Phys. Lett., 447, 168
\bibitem[(1981)]{Lee} Lee, L. C., \& Guest, J. A. 1981, J. Phys. B: At. Mol. Phys., 14, 3415
\bibitem[(2005)]{Lu} Lu, H.-C., Chen, H.-K., Cheng, B.-M., Kuo, Y.-P., \& Ogilvie, J. F. 2005, J. Phys. B: At. Mol. Opt. Phys., 38, 3693
\bibitem[(2008)]{Lu2} Lu, H.-C., Chen, H.-K., Cheng, B.-M., \& Ogilvie, J. F. 2008, Spectrochimica Acta Part A: Molecular and 
Biomolecular Spectroscopy, 71, 1485
\bibitem[(2006)]{Mason} Mason, N. J., Dawes, A., Holton, P. D., et al. 2006, Faraday Discussions, 133, 311
\bibitem[(1974)]{Monahan} Monahan, K. M., \& Walker, W. C. 1974, J. Chem. Phys., 61, 3886
\bibitem[(2005)]{Mota} Mota, R., Parafita, R., Giuliani, A., et al. 2005, Chem. Phys. Lett., 416, 152
\bibitem[(2011)]{Mumma} Mumma, M. J., \& Charnley, S. B. 2011, Annu. Rev. Astro. Astrophys., 49, 471
\bibitem[(2010)]{Caro1} Mu\~noz Caro, G. M., Jim\'enez-Escobar, A., Mart\'\i{}n-Gago, J.\'A., et al. 2010, A\&A, 522, A108
\bibitem[(1985)]{Nee} Nee, J. B., Suto, M., \& Lee, L. C. 1985, Chemical Physics, 98, 147
\bibitem[(2009)]{Oberg1} \"Oberg, K. I., Garrod, R. T., van Dishoeck, E. F., \& Linnartz, H. 2009a. A\&A, 504, 891
\bibitem[(2009)]{Oberg2} \"Oberg, K. I., Linnartz, H., Visser, R., \& van Dishoeck, E. F. 2009b. ApJ, 693, 1209
\bibitem[(2011)]{Oberg3} \"Oberg, K. I., Boogert, A. C. A., Pontoppidan, K. M., et al. 2011, ApJ, 740, 109
\bibitem[(1978)]{Okabe} Okabe, H. 1978, Photochemistry of small molecules, ed. John Wiley \& Sons, New York
\bibitem[(2008)]{Pontoppidan} Pontoppidan, K. M., Boogert, A. C. A., Fraser, H. J., et al. 2008, ApJ, 678, 1005
\bibitem[(1938)]{Price} Price, W. C., \& Simpson, D. M. 1938, Proc. Roy. Soc. (Lond.), A165, 272
\bibitem[(2012)]{Richey} Richey, C. R. \& Gerakines, P. A. 2012, ApJ, 759, 74
\bibitem[(2000)]{Vacuum} Samson, J. A. R., \& Ederer, D. L. 2000, Vacuum Ultaviolet Spectroscopy, ed. Elsevier Inc.
\bibitem[(1988)]{Sandford2} Sandford, S. A., Allamandola, L. J., Tielens, A. G. G. M., \& Valero, G. J. 1988, ApJ, 329, 498
\bibitem[(1990)]{Sandford3} Sandford, S. A., \& Allamandola, L. J. 1990, ApJ, 355, 357
\bibitem[(1996)]{Sandford} Sandford, S. A. 1996. Astronomical Society of the Pacific Conference Series, 97, 29
\bibitem[(1996)]{Schutte} Schutte, W. A. 1996. Molecules in Astrophysics: Probes and Processes, IAU symposium 178, 1
\bibitem[(2002)]{Smith} Smith, P. L., Rufus, L., Yoshino, K., \& Parkinson, W. H. 2002, NASA Laboratory Astrophysics Workshop, 
NASA/CP-2002-21186, 158
\bibitem[(1989)]{Sternberg} Sternberg, A. 1989, ApJ, 347, 863
\bibitem[(2007)]{Wu2} Wu, Y.-J., Lu, H.-C., Chen, H.-K., \& Cheng, B.-M. 2007, J. Chem. Phys., 127, 154311
\bibitem[(2012)]{Wu} Wu, Y.-J., Wu, C. Y. R., Chou S.-L., et al. 2012, ApJ, 746, 175
\bibitem[(1996)]{Yoshino} Yoshino, K., Esmond, J. R., Sun, Y., et al. 1996, J. Quant. Spectrosc. 
Radiat. Transfer, 55, 53
\end{thebibliography}
\end{document}